# Reprogrammable magnonic band structure of layered Permalloy/Cu/Permalloy nanowires

G. Gubbiotti,[1] X. Zhou,[2] Z. Haghshenasfard,[3] M. G. Cottam[3] and A.O. Adeyeye[2]

[1]Istituto Officina dei Materiali del CNR (CNR-IOM), Sede Secondaria di Perugia, c/o Dipartimento di Fisica e Geologia, Università di Perugia, I-06123 Perugia, Italy

[2]Information Storage Materials Laboratory, Department of Electrical and Computer Engineering, National University of Singapore, 117576 Singapore.

[3] Department of Physics and Astronomy, University of Western Ontario, London, Ontario N6A 3K7, Canada

ABSTRACT

Reprogrammability of magnonic band structure in layered Permalloy/Cu/Permalloy nanowires is demonstrated to depend on the relative orientation of the two layers magnetization. By using Brillouin light spectroscopy, we show that when the layers are aligned parallel two dispersive modes, with positive and negative group velocity, are observed while when the magnetic layers are aligned anti-parallel, only one dispersive mode, with positive group velocity, is detected. Our findings are successfully compared and interpreted in terms of a microscopic (Hamiltonian-based) method. An explanation for the observed behavior can be attributed to mode-mixing (or hybridization) effect when the two magnetic layers are aligned anti-parallel. This work opens the path to magnetic field-controlled reconfigurable magnonic crystals with multi-modal frequency transmission characteristics.

Magnonic crystals (MCs), which are magnetic metamaterials designed for the control of spin-wave (SW) propagation, are the subject of extensive investigation by the research community in the last few years [1,2,3,4,5] due to advances in lithographic techniques which enable the synthesis of highly ordered periodic nanostructures. Spectra of SW excitations in MCs are significantly different from those of uniform media and exhibit features such as magnonic band gaps, where SW propagation is entirely prohibited. Due to the wide tunability of their properties, MCs provide an excellent experimental testbed for the investigation of linear and nonlinear spin-wave dynamics.

Many of the efforts focus on reprogrammable magnonic crystals (MCs) whose dynamic response can be changed on demand. [6,7] In one-dimension (1D) the MCs with reprogrammable band structure typically have the form of periodic nanowire (NW) arrays with complex unit cells where the two subunits switch independently from one to another. Examples are arrays of two nanowires (NWs) of different widths or thicknesses [8,9,10] and NWs with the same cross-section size but formed using two different materials (bi-component MCs). [11,12,13,14] A specific advantage of 1D over 2D MCs is the fact that in the former case (e.g. utilizing longitudinally magnetized NWs) the complexities associated with the inhomogeneity of the internal magnetic field are absent.

So far, the reported SW propagation in MCs is limited to planar single-layer periodic magnetic structures where the band structure is controlled by dipolar coupling between the elements. In this work we investigate the SW band structure in dense arrays of layered Permalloy/Cu/Permalloy nanowires. The novelty of the proposed trilayer structure is that, due to the different layer thickness, one can control the relative magnetization orientation of the two layers in one and the same NW being either parallel (P) or anti-parallel (AP). Thus, by a proper magnetic field initialization, one can tune the overall magnonic band structure. We found that the dispersive modes are not the lowest frequency ones, as observed in dipolarly-coupled single layer NWs. All these features, including the field reprogrammable response, are peculiar to MCs, due to the non-volatile nature of magnetic materials, and do not find any counterpart in photonic [15] and phononic [16] crystals. Specific advantage of the proposed layered structure is that only one fabrication step is required for creating the patterns and that the thickness of the different layers can be controlled in an accurate manner down to the monolayer scale.

In this work, we investigate the SW band structure in 1D MCs constituted by dense arrays of Py/Cu/Py trilayer NWs where the two Py (Permalloy) NWs in a vertical arrangement have different thicknesses (10 and 30 nm). Four arrays of $Ni_{80}Fe_{20}$ (Permalloy, Py) NWs with trilayer Py($d_1$=10nm)/Cu(10 nm)/Py($d_2$=30 nm) structure were fabricated over an area of 100 × 100 μm$^2$ on an oxidized Si wafer substrate using high-resolution electron beam lithography, electron beam evaporation and lift-off processes. The trilayered NWs have a fixed width, $w$=280 nm, while their in-plane separation (edge-to-edge distance) ranges from $s$= 80 to 280 nm. The corresponding array period ($a$= $w$+$s$) and Brillouin zone (BZ) wave vector $\pi/a$ are summarized in the Table. The fabricated trilayer NWs were examined by scanning electron microscope and exhibit (see Fig. 1 for the sample with $s_1$=80 nm) uniform spacing and good edge definition.

Table I. Geometric parameters of the investigated nanowires arrays.

| Separation (nm) | Period (nm) | BZ wave vector ($10^7$ m$^{-1}$) |
|---|---|---|
| $s_1$=80 | $a_1$=360 | $\pi/a_1$= 0.87 |
| $s_2$=100 | $a_2$=380 | $\pi/a_2$= 0.82 |
| $s_3$=150 | $a_3$=430 | $\pi/a_3$= 0.73 |
| $s_4$=280 | $a_4$=560 | $\pi/a_4$= 0.56 |

The hysteresis (M-H) loops were measured in the longitudinal configuration using magneto-optical Kerr effect (MOKE) magnetometry. A dc magnetic field H, variable between -500 and +500 Oe, was applied in the sample plane, being parallel to the NWs length ($y$-direction) and perpendicular to the scattering plane ($x$-$z$ plane). The measured loop for the Py/Cu/Py NW arrays with separation $s_1$=80 nm displays a double-step switching behaviour, as shown in Fig. 1 (c). These two steps correspond to the reversal of the thin (10nm) NW, at a low field (-100 Oe), and the thick (30nm) NW at a high field (-160 Oe). In between, a state of anti-parallel alignment of the magnetization in the two Py layers is realized, similarly to what previously found in planar NW arrays with alternating width. [17] By contrast, arrays with separation from $s_2$ to $s_4$ exhibit almost identical square M-H loops with a coercivity of about 240 Oe.

The frequency dispersion (frequency vs wave vector) of thermally excited SWs were measured by Brillouin light scattering (BLS) spectroscopy. [18] The sample was mounted on a goniometer that allowed sample rotation around the field direction, i.e. to vary the incident angle of light, θ, between 0° (normal incidence) and 70°. The SW dispersion was mapped by changing the amplitude of the wave vector, $k$=(4π/λ)×sin(θ), parallel to the NWs width ($x$-direction), in the range from 0 to 1.9×$10^7$ rad/m, both in the the parallel (P) and antiparallel (AP) configurations. To create the P state, we first saturate the NWs along the positive $y$-direction at +500 Oe and then reduce to +150 Oe following the descending branch of the M-H loop. At this field the magnetization within the same NW layers and neighboring NWs are aligned along the field direction. The AP magnetization configuration within the same NW is attained by first saturating the array in the negative *y*-direction, then reducing the field to zero and finally reversing it until the +150 Oe value is reached.

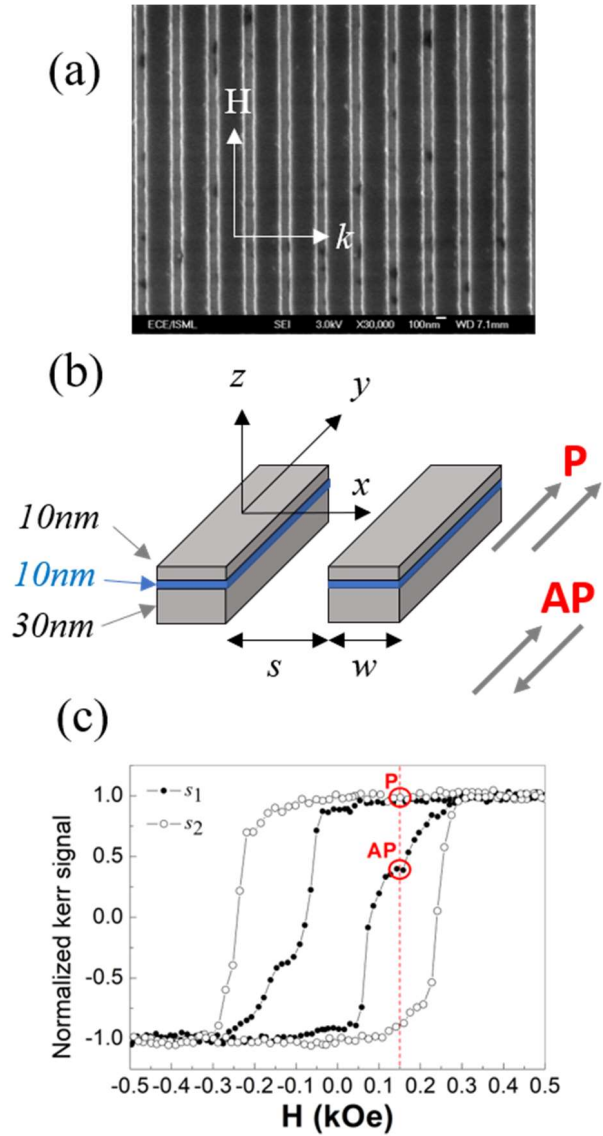

Fig. 1 (a) SEM image of the Py(10nm)/Cu(10nm)/Py(30nm) trilayer NW arrays with width of *w*=280 nm and edge-to-edge spacing of $s_1$=80 nm. The directions of the applied field (H) and wave vector (*k*) are also displayed. (b) Schematic drawing of the multilayer NWs shown together with the system of coordinate axes and the magnetization configuration in the P and AP states. (c) Measured MOKE loops for the NW arrays with $s_1$=80 nm and $s_2$=100 nm. The vertical red dashed line indicates the field value of +150 Oe which, following either the descending or the ascending branch of the hysteresis loop, corresponds to a parallel (P) or anti-parallel (AP) alignment between the two NWs magnetization for sample $s_1$.

The theory is based on a microscopic (or Hamiltonian-based) method that takes account of the exchange and magnetic dipolar interactions, as well as the magnetic anisotropy and effects of an external magnetic field. This approach was previously generalized from its earlier application to single magnetic nanoelements (such as nanowire stripes) [19] to MC arrays of simple NWs coupled by the dipolar fields across nonmagnetic spacers.[20] An important difference in this present work is the dipolar coupling across the spacer thickness $t$ in the out-of-plane (or $z$) direction as well as the dipolar coupling across the in-plane (or $x$) direction between the periodic elements of the MC (see Fig. 1). Extending Ref. [20], we assume that each element of the MC (the trilayer composite NW) is represented by spins arranged on a simple cubic lattice with an effective lattice constant chosen to be smaller than the exchange length of the magnetic material (which is about 5 nm for Py). The total Hamiltonian contains interaction terms describing the short-range exchange interactions within each magnetic component of a NW and the long-range dipole-dipole interactions within the stripes and across the spacers $s$ and $t$.

The calculations are a generalization of those reported in Ref. [20] since we now include the dipolar coupling in both the horizontal and vertical directions. Briefly, the translational symmetry along the NW length direction ($y$) and periodicity due to the MC lattice parameter $a$ in the $x$-direction are utilized to make a double Fourier transformation, thus introducing the wave number component $q$ and the Bloch wave number $k$, respectively. We assume that each layered NW has rectangular cross-section (in the $xz$ plane) and it is represented by spins arranged on a simple cubic lattice with an effective lattice constant $a$ chosen to be smaller than the exchange length of the magnetic material. The Hamiltonian can be expressed in terms of boson creation and annihilation operators, $a^+_{k,q,n}$ and $a_{k,q,n'}$, where $n$ and $n'$ (= 1, 2, …, $N$) label the $N$ spin sites in each cross section. The linearized SW spectrum is found from a bilinear term, which has the form

$$H^{(2)} = \sum_{k,q} \sum_{n,n'} \left[ A_{n,n'}(k,q) a^+_{k,q,n} a_{k,q,n'} + \left( B_{n,n'}(k,q) a_{k,q,n} a_{k,q,n'} + \text{h.c.} \right) \right], \quad (1)$$

where h.c. denotes Hermitian conjugate. The $A$ and $B$ coefficients depend on the exchange and dipolar sums, the applied field and the geometry, and they can be deduced as in Ref. [20]. Then, by making a canonical transformation to diagonalize Eq. (1) with respect to the $n$ and $n'$ labels, we may obtain the spectrum of discrete SW frequencies $\omega_\ell(k,q)$ for the MC, where $\ell$ labels the $N$ branches. The calculations include a small single-ion anisotropy (attributable to effects arising mainly at the lateral edges of the stripes), which plays a role in stabilizing the AP phase. As in Ref. [21] it is conveniently introduced in terms of an average effective anisotropy field which we have taken as $H_{an} = -80$ Oe in the present case.

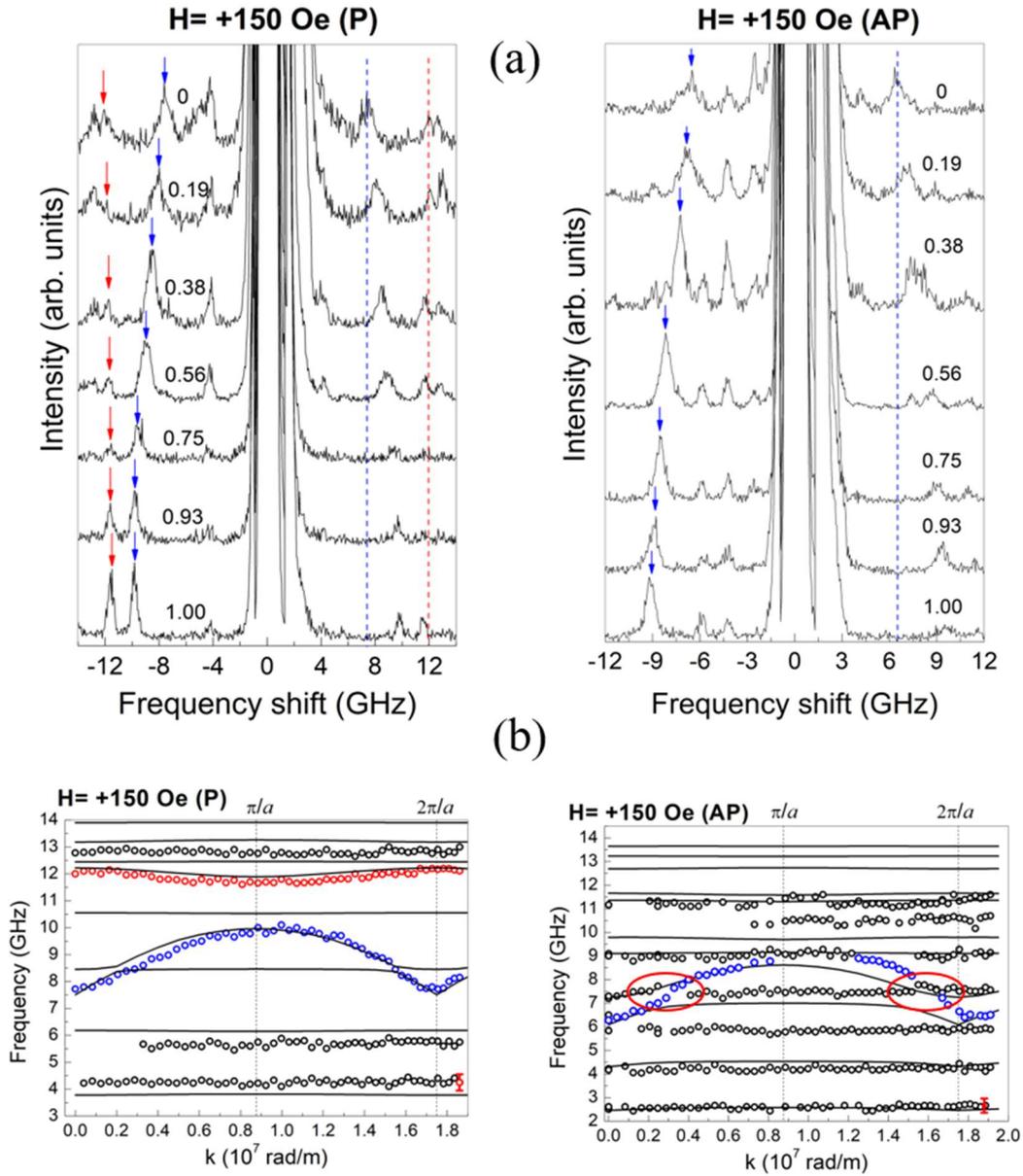

Fig. 2 (a) Measured BLS spectra as a function of the transferred wave vector $k$ (expressed in $\pi/a$) within the first BZ for NWs with spacing $s_1$=80 nm. The magnetic field H=150 Oe is applied along the NWs length and the orientation of magnetization within the same NW is parallel (P) and antiparallel (AP). Vertical blue and red dashed lines are guides to the eye to follow the frequency evolution of dispersive peaks. (b) Comparison between the measured (points) and calculated (lines) frequency dispersion for array $s_1$ in the P and AP states. A typical error bar is shown for the lowest frequency mode at $k=1.86\times10^7$ rad/m. Red ellipses indicate the wave vector region of modes hybridization while we plot in blue and red only modes with significant dispersion and largest peak intensity while all the others are marked in black.

Fig. 2 (a) shows a sequence of BLS spectra measured for the NW array with spacing $s_1$=80 nm at H=+150 Oe in both the P and AP states. All the spectra are characterized by the presence of several narrow peaks and only a few of them exhibit a sizeable frequency dependence on $k$. Some of them also are characterized by a significant variation of the peak intensity vs $k$. This can be explained by considering the modes we are observing in dense layered NWs are formed by the SW wave modes resonating through the NW width and coupled by dipolar dynamic interaction.[22] Therefore, modes with asymetric (symmetric) profiles with respect to the NW center width, are barely (amply) visible close to the center of the Brillouin zone ($k$=0) while they have large (low) intensity for $k$ values close to the edge of the first Brillouin zone (k=π/$a$). In addition, many previous studies have shown that the Stokes/anti-Stokes intensity ratio can be significantly affected by the magneto-optical coupling mechanisms[23] in the material and by the finite penetration depth of light which mainly probes the outermost NW layer.[24,25] Arrows in Fig. 2 (a) mark those peaks which exhibit the largest frequency variation thus indicating that collective SWs propagate through the array due to the dynamic dipolar coupling, while other peaks, either at lower or higher frequency with respect to this doublet of modes, are constant in frequency. Interestingly, in the P state the frequency difference between these two peaks, indicated by the red and blue arrows, decreases on increasing $k$ while in the AP state only one dispersive mode (blue arrow) is detected. A very good agreement between the calculated dispersion and the experimental one has been obtained by using the following parameters for Py obtained by fitting the dispersion of the continuous (unpatterned) Py(10)/Cu(10nm)/Py(30nm) films: namely, exchange stiffness $D$ = 30 T nm$^2$, the saturation magnetization M$_s$= 0.071 T, and $\gamma/2\pi$ = 29.3 GHz/T (with $\gamma$ being the gyromagnetic ratio).

Important information about the dispersion of SWs existing in the layered NW array can be derived by plotting the frequency vs the wave vector ($k$) values. In this sample, the peaks which exhibit the largest frequency variation show a dispersion that follows the Bloch's theorem with oscillating character and the occurrence of BZs due to the artificial periodicity of the NWs ($a$). These modes have a quite different width of the magnonic band (amplitude of the frequency oscillation) of 1.9 GHz for the blue peak and 0.4 GHz for the red one and are characterized by opposite dispersion slope at $k$ = 0 and $k$ = π/$a$. It is noteworthy, that the two dispersive modes are not the lowest frequency ones as in the case of single-layer dense NW arrays, where only the lowest frequency mode exhibits a significant dispersion and higher frequency modes are dispersionless.[26]

It is evident from Fig. 2(a) that the behavior of dispersive modes is quite different for the P and AP states with the same applied field H = + 150 Oe. The general appearance experimentally is characteristic of there being two dispersive modes in the P state with *opposite* dispersion (i.e. positive and negative group velocity), whereas in the AP state there seems to be one dispersive mode with positive group velocity. From the perspective of the theory, however, we see from Fig. 2 (b) in both cases the dispersive mode with positive group velocity is degenerate in frequency with dispersionless modes at about 8.5 GHz and 7 GHz in the P and AP states, respectively. The former is not detected in the measured spectra while the latter is visible and undergoes a large mode repulsion, due to hybridization at around k≈ 0.3×10$^7$ rad/m. It is important to note that switching from the P to AP configuration corresponds to a change in the interlayer dipolar energy and to an overall magnonic band shift downward by about 2 GHz.

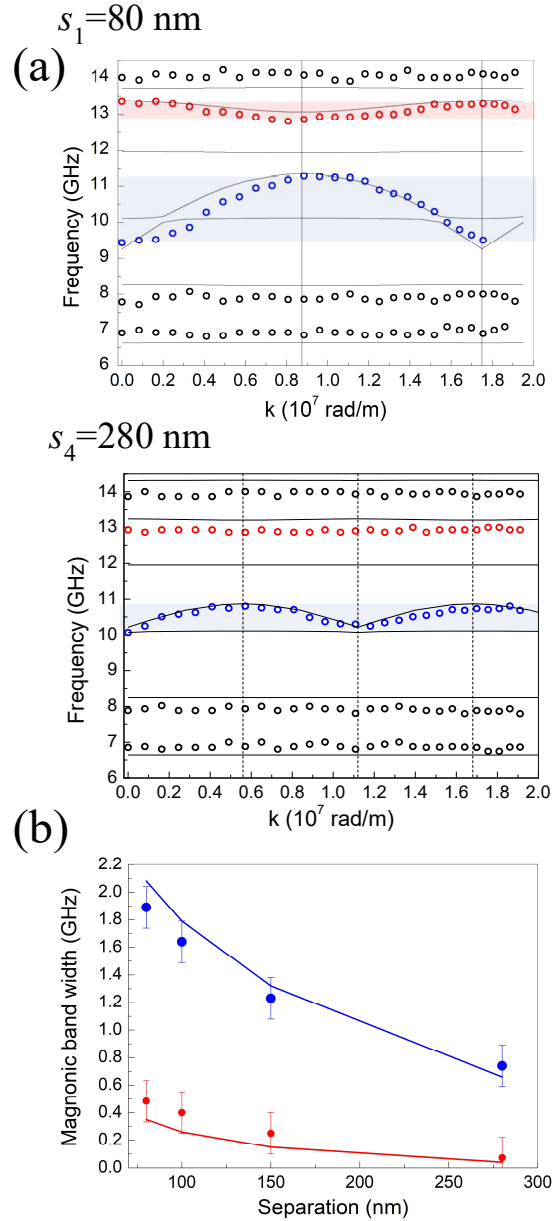

Fig. 3 (a) Comparison between the measured and calculated band structure for NW arrays with separations $s_1$ and $s_4$ for an applied field H=500 Oe. (b) Width of the magnonic band for the two most dispersive modes (marked by the red and blue dashed areas).

We next consider the case of NW arrays with different separation for an applied field H=500 Oe (i.e. NWs in the saturated P state) to investigate the role of lateral separation on the band structure. In Fig. 3(a) the comparison between the measured and calculated dispersion for arrays with smaller ($s_1$) and larger ($s_4$) separations are shown at H=500 Oe where all the NWs magnetization are aligned (P state). From inspection of the measured MOKE loops for NW arrays with separation $s_2$, $s_3$ and $s_4$ it is not possible to identify field regions of AP configuration. First of all, with separations larger than $s_1$, the SW dispersion can be mapped over the second BZ. The two lowest and the highest frequency modes are dispersionless and their frequency is independent of the NWs separation. By contrast, modes with frequencies between 9 and 13.5 GHz exhibit significant dispersion with a sizeable amplitude of the magnonic band which decreases with increasing $s$. For $s_1$ array, a significant reduction of the magnonic band width, from 2.5 to 1.9 GHz, is also observed for the most dispersive mode (blue points) on increasing the magnetic field value from H= 150 Oe (Fig. 2b) to 500 Oe (Fig. 3a) while it is almost constant for the less dispersive mode (red points). This is in agreement with previous findings where an increase of the magnetic field causes a reduction of the spin precession amplitude and consequently of the dynamic dipolar field which is responsible of the magnonic band width.[26,27] It is important to remark that by using the same set of magnetic parameters for the NW arrays, the measured frequency dispersion is well reproduced by our calculations (solid curves). Fig. 3 (b) summarizes theory and experiment for the evolution of the magnonic band width as a function of the NW separation. We see that the magnonic band is much wider for the lowest frequency dispersive mode than for the highest one and the band decreases monotonically in going from $s_1$ to $s_4$. This reduction is due to the steady decay (as $s$ is increased) of the dipolar coupling between adjacent NWs in the array.

In conclusion, we have demonstrated by experiments and theory the vertical reprogrammable band structure in Permalloy/Cu/Permalloy nanowires where the frequency position of stationary modes and the group velocity of the dispersive modes can be up- or down-shifted depending on the relative orientation of the magnetizations in the two Py layers. Additionally, we found that the widths of the magnonic band for two dispersive modes are controllable by the nanowires separation. In this respect, the investigated samples resemble spin-valve structures where the dynamic response can be changed on demand by reversing the magnetization orientation of one single layer. These peculiar properties go beyond the possibilities in photonics and plasmonics due to the intrinsic non-volatility of magnetization ground state configurations.

M.G.C. acknowledges support from the Natural Sciences and Engineering Research Council (NSERC) of Canada (grant RGPIN-2017-04429). A.O.A. was supported by the National Research